# Ab-initio insights into the pressure dependent physical properties and possible high-$T_c$ superconductivity in monoclinic and orthorhombic MgVH$_6$

*Md. Ashraful Alam[1,2], F. Parvin[2], S. H. Naqib[2]\**

[1]*Department of Physics, Mawlana Bhashani Science and Technology University, Santosh, Tangail 1902, Bangladesh*

[2]*Department of Physics, University of Rajshahi, Rajshahi 6205, Bangladesh*

*\*Corresponding author; Email: salehnaqib@yahoo.com*

**Abstract**

Here we have used the density functional theory (DFT) with the GGA-PBE approximation to investigate the structural, mechanical, electronic, hardness, thermal, superconductivity and optoelectronic properties under pressure for monoclinic ($P2_1/m$) and orthorhombic ($Pmn2_1$) structures of MgVH$_6$. We have studied optical properties of $P2_1/m$ phase at 0 GPa and $Pmn2_1$ phase at 100 GPa only (considering phase stability). Both of the phases of MgVH$_6$ are thermodynamically stable. $P2_1/m$ phase is mechanically stable but $Pmn2_1$ is mechanically unstable in our calculations for the pressures considered. Monoclinic ($P2_1/m$) is ductile in nature, on the other hand, orthorhombic ($Pmn2_1$) is brittle in nature at 100 GPa and becomes ductile for pressures in the range from 125 GPa to 200 GPa. Hardness calculations indicate superhard character of orthorhombic ($Pmn2_1$) structure at 100 GPa. The melting temperature of orthorhombic crystal is very high. This also agrees with the bulk modulus, Debye temperature, and hardness calculations. We have calculated theoretically the superconducting transition temperature $T_c$ at different pressures only for the orthorhombic ($Pmn2_1$) structure following a previous study. The estimated values of transition temperatures are within 104.7 K to 26.1 K in the pressure range from 100 GPa to 200 GPa. MgVH$_6$, in both the structures, are elastically and optically anisotropic.

## 1. Introduction

Anion of hydrogen is known as hydride. Usually this is a compound in which one or more hydrogen atom(s) is(are) present. According to Gibb a metal-to-hydrogen bond known as hydride while Blackledge defined hydride as a binary combination of hydrogen and a metal or metalloid [1]. On the other hand, according to the bonding nature, hydrides can be classified into four categories: (i) ionic hydride, (ii) covalent hydride, (iii) metallic hydride, and (iv) van der waals hydride [1]. Hydrogen can also be stored in three different phases; as gas, as liquid, and as solid. In solid hydrogen can be stored as chemical or physical combination with metals such as Li, Be, Na, Mg, B and Al, form a large variety of metal–hydrogen compounds and complex hydrides such as $Na_3AlH_6$, $Mg_2NiH_4$ [2]. There are various applications of hydrogen storage systems such as nickel-metal used as hydride batteries and metal hydrides ($M-H_n$) used as thermal storage for off-board storage [3]. Hydrogen-storage densities in metal hydrides (e.g., $MgH_2$) can be 6.5 H atoms/$cm^3$ while in gas it can be 0.99 H atoms/$cm^3$ and in liquid 4.2 H atoms/$cm^3$ [4]. $VH_2$ in the solid state contains 3.8 wt% hydrogen [5]. Metal hydride storage in solid is a better option than the gas or liquid storage system. At the same time, reversible hydrogen storage system with cyclic stability is very important. But the resistance of metal hydrides to impurities is one of the critical issues for on-board applications in order to maintain performance over the lifetime of the material [6,7].

Catalysis is also one of the serious factors in the improvement of hydrogen adsorption kinetics in metal hydride systems that enable dissociation of hydrogen molecules [8]. Palladium (Pd) is a good catalyst for hydrogen dissociation reaction [9]. Gennari et al., [10] used Pd, nickel (Ni), and germanium (Ge) as catalysts of hydrogenation kinetics. Hydrogen capacity can be increased up to 5.8 wt% using vanadium (V) as a catalyst [11].

Metal hydrides have attracted wide attention for not only as hydrogen (H) energy storage systems but also as potential high-temperature superconductors. For example, a number of H-rich binary compounds have shown superconductivity with very high superconducting transition temperature ($T_c$) under pressure e.g., 235 K for $CaH_6$ [12], 64 K for $GeH_4$ [13], 82 K for $LiH_6$ [14], 70 K for $KH_6$ [15], 38 K for $BeH_2$ [16] and so on [17-21]. $H_3S$ is a remarkably high superconductivity with a superconducting $T_c$ of 203 K at 200 GPa [22]. Transition metal hydrides are also interesting in the search for potential superconductors with high a $T_c$. For example, PdH at ambient pressure has a superconducting critical temperature of approximately 9 K with an unusual isotope effect [23]. Among the transition metals, V exhibits a transition temperature which increases from 5 K at 0 GPa to 17 K at 120 GPa [24-26].

Moreover, there are some predicted ternary hydrides with good superconducting properties under pressure such as ScCaH$_8$ and ScCaH$_{12}$ with the corresponding $T_c$ ~212 K and ~182 K, respectively, at 200 GPa [27]. ScYH$_6$ has $T_c$ ~ 32.110 K to 52.907 K in the pressure range 0-200 GPa [28] and H$_3$SXe with a $T_c$ of 89 K at 240 GPa [29]. The Mg based ternary hydrides are also investigated with predicted a MgGeH$_6$ $T_c$ up to 67 K at 200 GPa [30]. The predicted superconducting critical temperature of CaYH$_{12}$ 258 K at 200 GPa [31]. Besides, for LaSH$_6$ the estimated superconducting transition temperature is 35 K at 300 GPa [32], for MgSiH$_6$, $T_c$ ~ 63 K at 250 GPa [33], for MgScH$_6$, $T_c$ ~ 41 K at 100 GPa [34] and for MgVH$_6$, $T_c$ ~ 27.6 K at 150 GPa [35].

In general, Mg based hydrides have good-quality functional properties, such as heat-resistance, reversibility and recyclability. Among the Mg based hydrides MgH$_2$ is a promising material in solid-state due to its high storage capacity (7.6 wt% H$_2$), the availability, lower costs, crystal structure and superconductivity [35-38]. There are various Mg-H binary systems with predicted high temperature superconductivity at high pressures such as MgH$_6$ ($T_c$ = 260 K) at 300 GPa [39], MgH$_{12}$ ($T_c$ = 60 K) at 140 GPa [40]. Sun et al. [41] introduce extra electron in Mg-H system to tune the superconducting transition temperature. It was suggested by J. W. Arblaster that V-H systems should have excellent superconducting transition temperature at ambient pressure [42, 43]. Estimated $T_c$ of VH$_5$ is 24.5 K at 300 GPa [44] and for VH$_8$ it is 71.4 K at 200 GPa [45]. All these studies suggest that superconductivity of the binary Mg−H system may also be tuned by incorporating vanadium in the compound. Very recently, the structures of MgVH$_6$ were predicted by using the Crystal structure Analysis by Particle Swarm Optimization (CALYPSO) [46, 47]. Several phases such as *P*2$_1$/*m*, *C*2/*m*, *Pmn*2$_1$, *Cmc*2$_1$, *Cmcm*, and *P*4/*mmm* were predicted for ternary MgVH$_6$ compound at ambient and under different pressures [36]. Previous studies focused on the crystal structure, elastic properties, phase transition, electronic structures, and superconductivity to a limited extent of the ternary Mg-V hydrides at different pressures [36]. In this work we are inclined to address the structural, elastic, electronic, thermal, superconducting, and optical properties of MgVH$_6$ in two different structures at different pressures in detail using the density functional theory (DFT) based ab-initio methodology.

## 2. Computational scheme

For geometry optimization plane wave pseudopotential [48] method was implemented in the CASTEP code [49]. Generalized gradient approximation (GGA) with PBE functional [50] was used for the electronic exchange-correlation terms. Ultrasoft pseudopotential was used for the calculations of electron-ion interactions [51]. Plane-wave cutoff of energy 380 eV was used throughout the calculations and the BFGS algorithm was used to minimize the total energy and internal forces [52]. Size of the Monkhorst–Pack grid [53] was 4×8×5 for $P2_1/m$ phase and 5×5×5 for $Pmn2_1$ phase for sampling of the first Brillouin zone (BZ). The quality of convergence tolerance for optimizing the geometry was set as: difference in total energy $5\times10^{-6}$ eV/atom, maximum force 0.01 eV/Å, maximum stress 0.02 GPa, maximum displacement 0.001 Å, and self-consistent field tolerance within $5\times10^{-4}$ eV/atom. The valence electron configurations were as follows: Mg[$2p^63s^2$], V[$3s^23p^63d^34s^2$], H[$1s^1$]. The elastic constants were determined using the stress-strain module in CASTEP. The optical constants were calculated from the matrix element of optical transition of electrons from the valence band to the conduction band. Thermophysical parameters were computed from the elastic constants and moduli.

## 3. Results and analysis

### 3.1 *Structure and stability*

The crystal structures of monoclinic and orthorhombic crystal system belonging to the space groups $P2_1/m$ and $Pmn2_1$, respectively, are shown in Fig. 1. In Fig. 1 we represent the crystal structure at 0 GPa and 100 GPa of (a) monoclinic and (b) orthorhombic structures of MgVH$_6$, respectively. For both the monoclinic and orthorhombic crystal structures the unit cell contains two formula units with sixteen atoms in total. The calculated lattice parameters are listed in Table 1 with the previous values for comparison. The optimized lattice parameters are consistent with the previous work for both of the structure. In this work, we have calculated the cohesive energy per atom using the equation [54-56]

$$E_{Coh} = \frac{E_{Mg} + E_V + 6E_H - E_{MgVH_6}}{8} \qquad (1)$$

Where, $E_{MgVH_6}$ is total energy per formula unit of MgVH$_6$ and $E_{Mg}$, $E_V$ and $E_H$ are total energy of single Mg, V and H atoms, respectively. From Table 1 it is found that the values of cohesive energy per atom are positive for MgVH$_6$, which indicate that two phases $P2_1/m$ and $Pmn2_1$ of MgVH$_6$ are thermodynamically stable. The values of lattice parameters *a* and *c* decrease at a faster rate than *b* with pressure indicating that *a* and *c* directions of the $P2_1/m$ crystal are easily compressible than the *b* direction under pressure. For the $Pmn2_1$ structure the pressure dependences of a, b, and c are similar with significantly different behavior occurring at pressures above 170 GPa. This is an indication of structural instability.

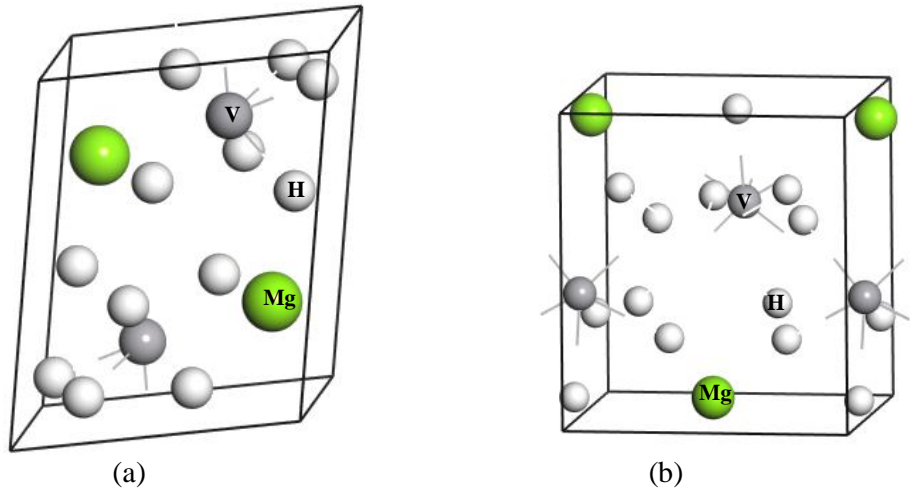

Figure 1: Crystal structures of (a) $P2_1/m$ and (b) $Pmn2_1$ MgVH$_6$.

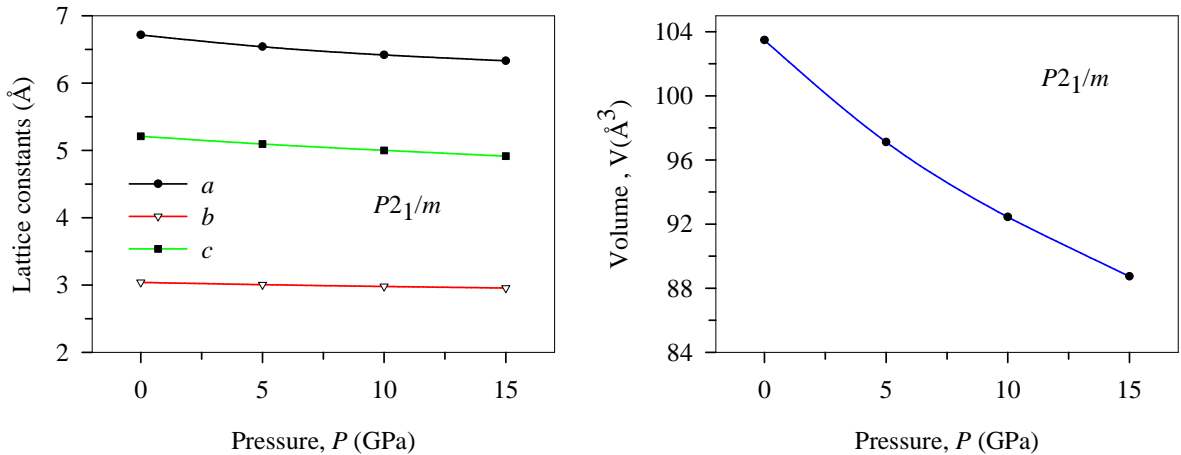

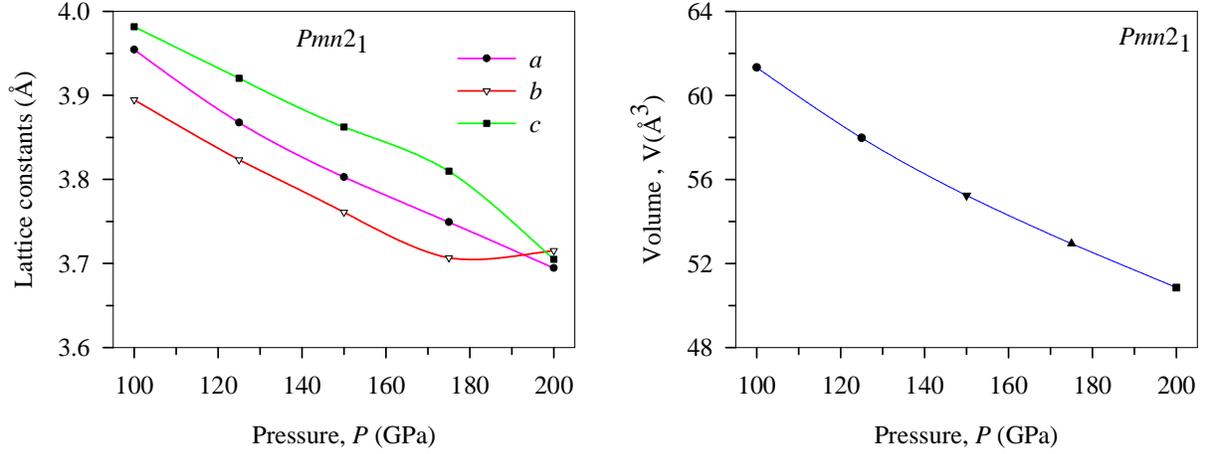

Figure 2: Lattice parameters under pressure of MgVH$_6$ in two different structures.

Table 1. The calculated lattice parameters $a$ (Å), $b$ (Å), $c$ (Å), cell volume $V$ (Å$^3$), and cohesive energy $E_{coh}$ (eV/atom) of MgVH$_6$ in different structures.

| Space group | Crystal structure | Pressure $P$ (GPa) | $a$ | $b$ | $c$ | $V$ | $E_{coh}$ | Ref. |
|---|---|---|---|---|---|---|---|---|
| $P2_1/m$ | Monoclinic | 0 | 6.65870 | 3.02040 | 5.13840 | - | - | [36] |
| | | 0 | 6.71398 | 3.03794 | 5.20925 | 103.47 | 3.85 | [This] |
| | | 5 | 6.53858 | 3.00484 | 5.09337 | 97.11 | 3.84 | |
| | | 10 | 6.41669 | 2.97763 | 4.99983 | 92.44 | 3.83 | |
| | | 15 | 6.32944 | 2.95462 | 4.91356 | 88.73 | 3.81 | |
| | | 25 | 6.19790 | 2.89270 | 4.75360 | - | - | [36] |
| $Pmn2_1$ | Orthorhombic | 100 | 3.9544 | 3.8947 | 3.9817 | 61.32 | 3.28 | [This] |
| | | 125 | 3.8676 | 3.8233 | 3.9204 | 57.97 | 3.14 | |
| | | 150 | 3.8028 | 3.7608 | 3.8622 | 55.24 | 2.99 | |
| | | 150 | 3.8108 | 3.7582 | 3.8433 | - | - | [36] |
| | | 175 | 3.7490 | 3.7066 | 3.8099 | 52.94 | 2.85 | [This] |
| | | 200 | 3.6945 | 3.7154 | 3.7049 | 50.85 | 2.69 | |
| | | 200 | 3.6943 | 3.6527 | 3.7694 | - | | [36] |

## 3.2 Elastic properties

### 3.2.1 Single crystal elastic constants

For the monoclinic crystal structure, there are thirteen independent elastic constants: $C_{11}$, $C_{22}$, $C_{33}$, $C_{44}$, $C_{55}$, $C_{66}$, $C_{12}$, $C_{13}$, $C_{15}$, $C_{23}$, $C_{25}$, $C_{35}$ and $C_{46}$. For the orthorhombic crystal structure there are nine independent elastic constants: $C_{11}$, $C_{22}$, $C_{33}$, $C_{44}$, $C_{55}$, $C_{66}$, $C_{12}$, $C_{13}$ and $C_{23}$. The computed independent elastic constants are listed in Table 2. From Table 2 it is observed that most of the elastic constants increase with increasing pressure for both of the phases except $C_{15}$, $C_{25}$, $C_{35}$, $C_{46}$ for the monoclinic and $C_{22}$, $C_{44}$ for the orthorhombic structure. The negative values of elastic constants imply that the optimized geometries are not completely relaxed and internal strains are present within the crystals. The largest values of the elastic constants are found for $C_{11}$, $C_{22}$, $C_{33}$ for both of the structures indicating that deformation resistances along the *a*-, *b*- and *c*-axis are very strong. According to Liu, Q. J. et al. [57] the mechanical stability conditions of monoclinic and orthorhombic crystals under pressure are as follows:

For the monoclinic structure:

$$C_{11} - P > 0, C_{22} - P > 0, C_{33} - P > 0, C_{44} - P > 0, C_{55} - P > 0, C_{66} - P > 0 \tag{2}$$

$$C_{11} + C_{22} + C_{33} + 2C_{12} + 2C_{13} + 2C_{23} + 3P > 0 \tag{3}$$

$$(C_{11} - P)(C_{33} - P) > C_{35}^2, (C_{44} - P)(C_{66} - P) > C_{46}^2, (C_{22} + C_{33} - 2C_{23} - 4P) > 0 \tag{4}$$

$$(C_{22} - P)[(C_{33} - P)(C_{55} - P) - C_{35}^2] + 2(C_{23} - P)C_{25}C_{35}$$
$$> (C_{23} + P)^2(C_{55} - P) + C_{25}^2(C_{33} - P) \tag{5}$$

$$\begin{aligned}
2C_{15}C_{25}&[(C_{33} - P)(C_{12} + P) - (C_{13} + P)(C_{23} + P)] \\
&+ 2C_{15}C_{35}[(C_{22} - P)(C_{13} + P) - (C_{12} + P)(C_{23} + P)] \\
&+ 2C_{25}C_{35}[(C_{11} - P)(C_{12} + P) - (C_{12} + P)(C_{13} + P)] \\
&- C_{15}^2[(C_{22} - P)(C_{33} - P) - (C_{23} + P)^2] - C_{25}^2[(C_{11} - P)(C_{33} - P) - (C_{13} + P)^2] \\
&- C_{35}^2[(C_{11} - P)(C_{22} - P) - (C_{12} + P)^2] \\
&+ (C_{55} - P)[(C_{13} - P)(C_{22} - P)(C_{33} - P) - (C_{11} - P)(C_{23} + P)^2 \\
&- (C_{22} - P)(C_{13} + P)^2 - (C_{33} - P)(C_{12} + P)^2 + 2(C_{12} + P)(C_{13} + P)(C_{23} + P)] \\
&> 0
\end{aligned} \tag{6}$$

For the Orthorhombic structure:

$$C_{11} - P > 0, C_{22} - P > 0, C_{33} - P > 0, C_{44} - P > 0, C_{55} - P > 0, C_{66} - P > 0 \tag{7}$$

$$, (C_{11} + C_{22} - 2C_{12} - 4P) > 0, (C_{11} + C_{33} - 2C_{13} - 4P) > 0, (C_{22} + C_{33} - 2C_{23} - 4P)$$
$$> 0, (C_{11} + C_{22} + C_{33} + 2C_{12} + 2C_{13} + 2C_{23} + 3P) > 0 \tag{8}$$

Our calculated result shows that $P2_1/m$ phase satisfies all conditions under pressure so $P2_1/m$ phase is mechanically stable under pressures up to 15 GPa. On the other hand, $Pmn2_1$ phase is mechanically unstable because it does not satisfy all the stability conditions.

Table 2. Calculated single crystal elastic constants $C_{ij}$ (GPa) for MgVH$_6$.

| $C_{ij}$ | Monoclinic ($P2_1/m$) | | | | Orthorhombic ($Pmn2_1$) | | | | | Ref. |
| --- | --- | --- | --- | --- | --- | --- | --- | --- | --- | --- |
| | Pressure, $P$ (GPa) | | | | Pressure, $P$ (GPa) | | | | | |
| | 0 | 5 | 10 | 15 | 100 | 125 | 150 | 175 | 200 | |
| $C_{12}$ | 34.05 | 53.51 | 57.08 | 87.86 | 319.65 | 406.81 | 432.99 | 488.64 | 557.85 | [This] |
| | | | | | | | 382.26 | | 436.48 | [36] |
| $C_{13}$ | 50.86 | 73.44 | 94.24 | 114.39 | 375.27 | 413.40 | 483.18 | 533.54 | 597.46 | [This] |
| | | | | | | | 321.99 | | 418.99 | [36] |
| $C_{15}$ | 2.89 | 8.65 | 15.04 | 0.04 | | | | | | [This] |
| $C_{23}$ | 51.00 | 73.01 | 75.06 | 105.00 | 335.68 | 379.68 | 449.80 | 503.33 | 574.03 | [This] |
|          |       |       |       |        |        |        | 376.62 |        | 436.01 | [36]   |
| $C_{25}$ | 0.60  | 0.85  | 2.36  | -15.32 |        |        |        |        |        | [This] |
| $C_{35}$ | -0.21 | 13.14 | 27.48 | 20.67  |        |        |        |        |        |        |
| $C_{46}$ | -4.89 | -4.60 | -2.92 | -0.56  |        |        |        |        |        |        |
| $C_{11}$ | 100.91| 129.36| 160.96| 187.27 | 543.70 | 624.08 | 757.90 | 860.94 | 955.83 | [This] |
|          |       |       |       |        |        |        | 513.36 |        | 695.30 | [36]   |
| $C_{22}$ | 202.10| 249.65| 258.26| 322.00 | 551.50 | 633.09 | 728.24 | 838.94 | 890.37 | [This] |
|          |       |       |       |        |        |        | 503.34 |        | 695.30 | [36]   |
| $C_{33}$ | 100.28| 131.18| 155.87| 174.87 | 614.29 | 649.56 | 767.22 | 851.96 | 949.79 | [This] |
|          |       |       |       |        |        |        | 584.92 |        | 714.71 | [36]   |
| $C_{44}$ | 46.15 | 55.23 | 62.35 | 66.35  | 159.34 | 179.50 | 197.45 | 211.80 | 223.33 | [This] |
|          |       |       |       |        |        |        | 200.50 |        | 98.88  | [36]   |
| $C_{55}$ | 28.85 | 35.74 | 41.30 | 47.86  | -41.93 | -98.20 | 103.17 | 157.98 | 228.81 | [This] |
|          |       |       |       |        |        |        | 249.05 |        | 98.88  | [36]   |
| $C_{66}$ | 37.11 | 43.89 | 49.76 | 54.27  | 145.10 | 163.25 | 181.49 | 197.00 | 212.21 | [This] |
|          |       |       |       |        |        |        | 187.95 |        | 211.10 | [36]   |

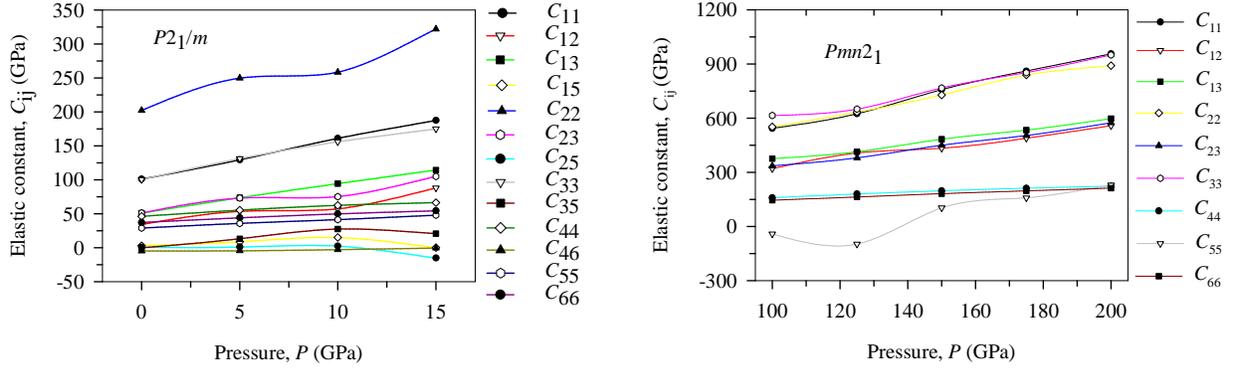

Figure 3: Single crystal elastic constants $C_{ij}$ (GPa) under pressure of MgVH$_6$ compounds with different structures.

From Fig. 3 it is observed that the elastic constant $C_{55}$ of MgVH$_6$ in the orthorhombic phase shows strongly nonmonotonic variation with pressure in the range from 100 GPa to 160 GPa. This observation might be linked with a possible structural instability as seen in Fig. 2 (for $Pmn2_1$ structure) at somewhat higher pressure.

### 3.2.2 Polycrystalline elastic properties

The polycrystalline elastic parameters such as Young's modulus, bulk modulus, shear modulus Poisson's ratio, Pugh's ratio, and machinability index are important indicators to evaluate the mechanical performance of a solid. As we know, Young's modulus represents the ability of a solid to counter uniaxial tension, shear modulus represents the resistance to shape changing plastic deformation, and bulk modulus can be used to characterize the resistance to change in the volume. The value of Young's modulus $Y$, shear modulus $G$, bulk modulus $B$ are calculated of monoclinic and orthorhombic structures for MgVH$_6$ by using the Viogt-Reuss-Hill (VRH) approximation [58-60].

Table 3. Calculated polycrystalline bulk modulus $B$ (GPa), shear modulus $G$ (GPa), Young's modulus $Y$ (GPa), Pugh's ratio $G/B$, Poisson's ratio $v$, and machinability index $\mu_m$ for MgVH$_6$.

| Space group | Crystal structure | Pressure P (GPa) | $B$ | $G$ | $Y$ | $G/B$ | $v$ | $\mu_m$ | Ref. |
|---|---|---|---|---|---|---|---|---|---|
| $P2_1/m$ | Monoclinic | 0 | 72.45 | 37.73 | 96.44 | 0.52 | 0.28 | 1.57 | [This] |
| | | 5 | 96.95 | 44.47 | 115.71 | 0.46 | 0.30 | 1.76 | |
| | | 10 | 109.51 | 49.69 | 129.48 | 0.45 | 0.30 | 1.76 | |
| | | 15 | 140.82 | 53.56 | 142.60 | 0.38 | 0.33 | 2.12 | |
| $Pmn2_1$ | Orthorhombic | 100 | 417.19 | 391.84 | 895.25 | 0.94 | 0.142 | 2.62 | [This] |
| | | 125 | 478.41 | 181.68 | 483.79 | 0.38 | 0.331 | 2.67 | |
| | | 150 | 553.16 | 151.63 | 416.81 | 0.27 | 0.374 | 2.80 | |
| | | 175 | 622.22 | 180.58 | 493.95 | 0.29 | 0.368 | 2.94 | |
| | | 200 | 694.03 | 202.65 | 554.02 | 0.29 | 0.367 | 3.11 | |

The calculated values of polycrystalline elastic moduli, Pugh's ratio $G/B$, Poisson's ratio $v$, machinability index $\mu_m$ are listed in Table 3. It can be seen from Table 3 and Fig. 4 that the values of Young's modulus, shear modulus, bulk modulus of monoclinic MgVH$_6$ structure increase with pressure which indicates the resistance to plastic deformation and uniaxial tension is enhanced up to 15 GPa. For orthogonal structure of MgVH$_6$ only the bulk modulus increases with pressure and the values of shear and Young's modulus are gradually decrease up to 150 GPa. This behavior is anomalous and once again points towards a tendency of structural instability. Sudden increase in the Poisson's ratio from 0.142 (at 100 GPa) to 0.331 (at 125 GPa) is another indication that there is a drastic change in the bonding characteristics in this pressure range. Above 150 GPa, these values increase which indicates the resistance to plastic deformation and uniaxial tension is enhanced from 150 GPa to 200 GPa for the orthorhombic MgVH$_6$.

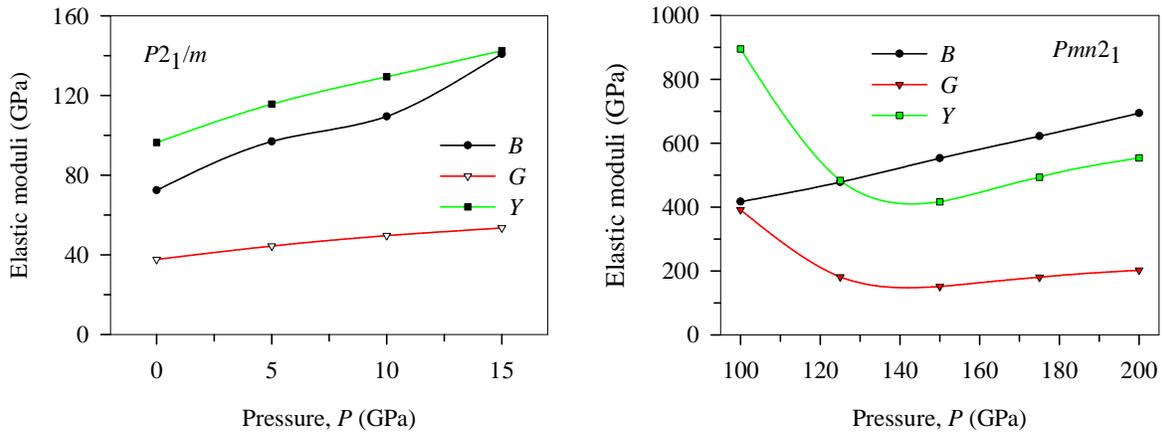

Figure 4: Polycrystalline elastic constant under pressure of MgVH$_6$ in the monoclinic and orthorhombic structures.

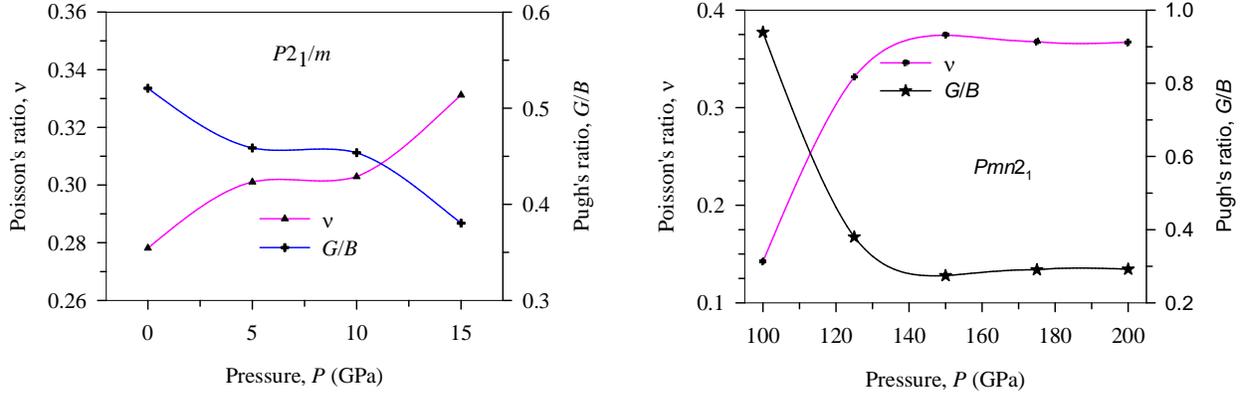

Figure 5: Poisson's ratio and Pugh's ratio under pressure of MgVH$_6$ in the monoclinic and orthorhombic structures.

The value of Poisson's ratio is obtained using the equation, $\nu = (3B - 2G)/(3B + G)$. This value indicates the ductile and brittle nature of a compound. According to Frantsevich [61], if $\nu >$ 0.26 the solid should exhibit ductile nature otherwise it should be brittle. For the monoclinic structure (*P2$_1$/m*), the values of $\nu$ are greater than 0.26 which indicates the ductile nature of *P2$_1$/m* phase of MgVH$_6$ for the pressures considered. Poisson's ratio and hence the degree of ductility increases with increasing pressure. On the other hand, for the *Pmn2$_1$* phase, the value of $\nu$ is less than 0.26 at 100 GPa, and from 125 GPa to 200 GPa, $\nu$ is larger than 0.26. This indicates the brittle nature at 100 GPa and ductile nature in the pressure range from 125 GPa to 200 GPa. The Pugh's ratio (*G/B*) [63] is another parameter used to define ductile and brittle nature of a compound [62]. *G/B* > 0.57 indicates the brittle nature and otherwise ductility. From Table 3 it is observed that the Pugh's ratio indicates the same nature as the Poisson's ratio. Fig. 5 shows an anticorrelation between the Poisson's ratio and the Pugh's ratio.

*3.3 Elastic anisotropy*

Study of the elastic anisotropy is important to understand the direction dependent bonding characteristics and mechanical properties of solids. The shear anisotropy factors [63] obtained from our calculations are given in Table 4. These factors $A_1$, $A_2$, and $A_3$ must be one for an isotropic crystal, while any value except unity is a measure of the degree of elastic anisotropy possessed by the crystal. The factors $A_1$, $A_2$, and $A_3$ are computed from the expressions given follows:

$$A_1 = \frac{4C_{44}}{C_{11} + C_{33} - 2C_{13}}, A_2 = \frac{4C_{55}}{C_{22} + C_{33} - C_{23}}, A_3 = \frac{4C_{66}}{C_{11} + C_{22} - 2C_{12}} \qquad (9)$$

Furthermore, $A^B$ and $A^G$ are the percentage anisotropies in compressibility and shear, respectively and $A^U$ is the universal anisotropic factor. The zero values of $A^B$, $A^G$ and $A^U$ represent elastical isotropy of a crystal and non-zero values represent anisotropy [64, 65]. The values of $A^U$, $A^B$, and $A^G$ are also listed in Table 4. These anisotropy indices are calculated using the following equations:

$$A^B = \frac{B_V - B_R}{B_V + B_R}, A^G = \frac{G_V - G_R}{G_V + G_R} \text{ and } A^U = 5\frac{G_V}{G_R} + \frac{B_V}{B_R} + 6 \geq 0 \tag{10}$$

Table 4. Calculated elastic anisotropic factors of $MgVH_6$.

| Space group | Crystal structure | Pressure P (GPa) | $A_1$ | $A_2$ | $A_3$ | $A^U$ | $A^B$ | $A^G$ | Ref. |
|---|---|---|---|---|---|---|---|---|---|
| $P2_1/m$ | Monoclinic | 0 | 1.86 | 0.58 | 0.63 | 0.79 | 0.04 | 0.14 | [This] |
| | | 5 | 1.94 | 0.61 | 0.65 | 0.86 | 0.04 | 0.15 | |
| | | 10 | 1.94 | 0.63 | 0.65 | 1.02 | 0.04 | 0.19 | |
| | | 15 | 1.99 | 0.67 | 0.65 | 1.14 | 0.02 | 0.22 | |
| $Pmn2_1$ | Orthorhombic | 100 | 1.56 | -0.34 | 1.27 | -4.28 | 0.004 | -0.75 | [This] |
| | | 125 | 1.61 | -0.75 | 1.47 | -3.20 | 0.000 | -0.47 | |
| | | 150 | 1.41 | 0.69 | 1.17 | 0.27 | 0.001 | 0.03 | |
| | | 175 | 1.31 | 0.92 | 1.09 | 0.07 | 0.001 | 0.01 | |
| | | 200 | 1.26 | 1.32 | 1.16 | 0.07 | 0.001 | 0.01 | |

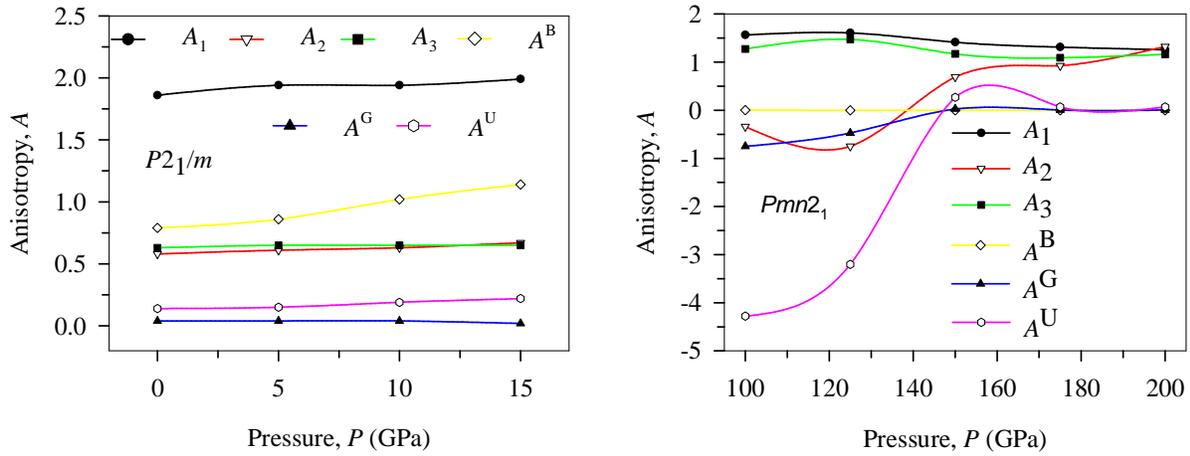

Figure 6: Anisotropy factors under pressure of $MgVH_6$ in the monoclinic and orthorhombic structures.

Table 4 (and Fig. 6) shows clearly that $MgVH_6$ in both monoclinic and orthorhombic structures are elastically anisotropic. These anisotropies are clear indications that atomic bonding strengths in different directions in the crystals are different. The negative values of $A^U$ in the orthorhombic structure at 100 GPa and 125 GPa are indicative of structural instability.

*3.4 Acoustic sound velocities*

The sound velocities in a crystal are useful thermophysical parameter. They are closely related to the crystal stiffness and crystal density and determine the Debye temperature and thermal conductivity to a large extent. Crystalline solids support both longitudinal and transverse modes of propagation of acoustic disturbances. The calculated values of Sound velocities ($v_t$, $v_l$ and $v_m$) under pressure of MgVH$_6$ are listed in Table 5 using the following equations [66-68]:

$$v_t = \sqrt{\frac{G}{\rho}}, \ v_l = \sqrt{\frac{3B + 4G}{3\rho}} \text{ and } v_m = \left[\frac{1}{3}\left(\frac{2}{v_t^3} + \frac{1}{v_l^3}\right)\right]^{-\frac{1}{3}} \quad (11)$$

The computed values at different pressures are enlisted in Table 5 and shown in Fig. 7.

Table 5. Calculated density $\rho$ (gm/cm$^3$), ), transverse sound velocities $v_t$ ( km/s), longitudinal sound velocities $v_l$ ( km/s) and average sound velocities $v_m$ ( km/s) of MgVH$_6$.

| Space group | Crystal structure | Pressure P (GPa) | $\rho$ | $v_t$ | $v_l$ | $v_m$ | Ref. |
|---|---|---|---|---|---|---|---|
| $P2_1/m$ | Monoclinic | 0 | 2.609 | 3.8026 | 6.8591 | 4.2359 | [This] |
|  |  | 5 | 2.780 | 3.9997 | 7.4973 | 4.4682 |  |
|  |  | 10 | 2.920 | 4.1249 | 7.7581 | 4.6092 |  |
|  |  | 15 | 3.042 | 4.1956 | 8.3520 | 4.7054 |  |
| $Pmn2_1$ | Orthorhombic | 100 | 1.597 | 15.6657 | 24.2591 | 17.1933 | [This] |
|  |  | 125 | 1.689 | 10.3713 | 20.6561 | 11.6318 |  |
|  |  | 150 | 1.773 | 9.2489 | 20.6426 | 10.4332 |  |
|  |  | 175 | 1.849 | 9.8814 | 21.6018 | 11.1365 |  |
|  |  | 200 | 1.925 | 10.2594 | 22.3790 | 11.5613 |  |

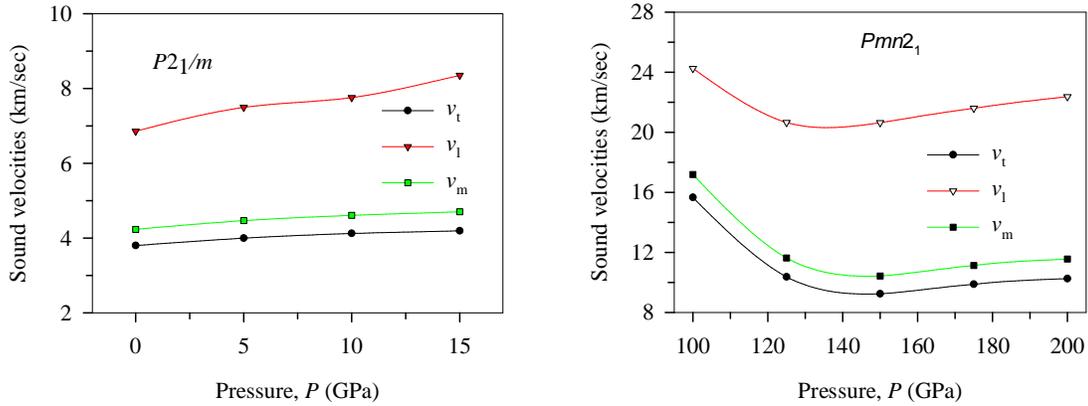

Figure 7: Sound velocities ($v_t$, $v_l$ and $v_m$) under pressure of MgVH$_6$ in the monoclinic and orthorhombic structures.

From Fig. 7 [cf. Table 5] it is observed that for the $P2_1/m$ phase, all the sound velocities increase monotonously with pressure. For the $Pmn2_1$ phase, $v_t$ decreases in the pressure range from 100 GPa to 125 GPa and increases after 125 GPa with increase in the pressure. Sound velocity is greatly enhanced in the orthorhombic phase. This is a consequence of the greatly enhanced stiffness of the MgVH$_6$ compound in the orthorhombic structure.

*3.5 Hardness of MgVH$_6$*

Hardness of a solid is used to assess both elastic and plastic behavior of the material under mechanical stress. This particular parameter determines the average bonding strength and stiffness of a solid. A variety of approaches are available to compute the hardness of a system. In this study we have calculated the hardness using the formalisms developed by Teter et al. [69], Tian et al. [70], and Chen et al. [71]. We have also calculated the micro hardness [72] of MgVH$_6$ at different pressures in the monoclinic and orthorhombic structures. The calculated values of hardness are presented in Table 6. The pressure dependent hardness values are also illustrated in Fig. 8.

Table 6. Hardness of MgVH$_6$ in the monoclinic and orthorhombic structures.

| Space group | Crystal structure | Pressure P (GPa) | Hardness (GPa) | | | | Ref. |
|---|---|---|---|---|---|---|---|
| | | | $H_{Teter}$ | $H_{Tian}$ | $H_{Chen}$ | $H_{micro}$ | |
| $P2_1/m$ | Monoclinic | 0 | 5.70 | 4.92 | 4.79 | 5.58 | [This] |
| | | 5 | 6.71 | 4.64 | 4.40 | 5.90 | |
| | | 10 | 7.50 | 4.95 | 4.79 | 6.53 | |
| | | 15 | 8.09 | 4.10 | 3.62 | 6.03 | |
| $Pmn2_1$ | Orthorhombic | 100 | 59.17 | 57.87 | 58.12 | 93.43 | [This] |
| | | 125 | 27.43 | 9.71 | 10.51 | 20.41 | |
| | | 150 | 22.90 | 5.47 | 5.30 | 12.69 | |
| | | 175 | 27.27 | 6.69 | 6.83 | 15.93 | |
| | | 200 | 30.60 | 7.32 | 7.59 | 17.97 | |

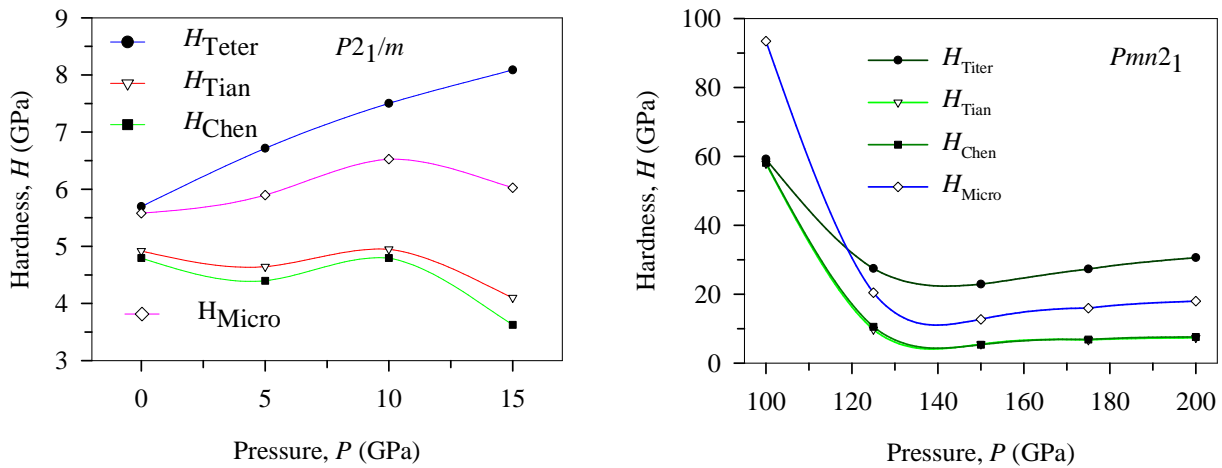

Figure 8: Hardness under pressure of MgVH$_6$ in the monoclinic and orthorhombic structures.

Fig. 8 shows that different formalisms result in different values of hardness. In the monoclinic phase Teter formalism yields the highest hardness. Overall the hardness of MgVH$_6$ in the monoclinic structure is moderate. At 100 GPa, MgVH$_6$ in the orthorhombic phase exhibits anomalously high hardness which falls sharply with increasing pressure. This is an indication of a drastic modification of pressure induced bonding strength in the orthorhombic phase. At high enough pressures both the structures are quite soft.

*3.6 Thermodynamic properties*

*Debye temperature*: The Debye temperature $\theta_D$ of a system is closely connected to many physical properties of solids such as specific heat, melting temperature, thermal conductivity, hardness, elastic constants, bonding strength, and sound velocity. The knowledge of $\theta_D$ provides information about the electron-phonon coupling and Cooper pairing mechanism of superconductivity. The Debye temperature calculated from the elastic constants is considered to be similar to that acquired from the specific heat measurements. Using the average sound velocity, the Debye temperature can be calculated by using the Anderson method [73] as follows:

$$\theta_D = \frac{h}{k_B}\left[\left(\frac{3n}{4\pi}\right)\frac{N_A\rho}{M}\right]^{1/3} v_m \tag{12}$$

where, $h$ is Planck's constant, $k_B$ is the Boltzmann's constant, $V$ is the volume of unit cell, $n$ is the number of atoms within a unit cell, and $v_m$ is the average sound velocity. The computed results are disclosed in Table 7.

*Minimum thermal conductivity*: At temperatures above $\theta_D$ the thermal conductivity of a solid attains a minimum value known as the minimum thermal conductivity, $\kappa_{min}$. The calculated values of the minimum thermal conductivity are derived from the relation given by $k_{min} = k_B v_m (V_a)^{-2/3}$ [74] and presented in Table 7.

*Grüneisen parameter*: The Grüneisen parameter $\gamma$ is an important thermophysical quantity that links the vibrational properties with the structural ones. It is related to the expansion coefficient, bulk modulus, specific heat, and electron-phonon coupling in solids. The normal thermal expansion of solids due to anharmonicity of interatomic forces is understood from the Grüneisen constant as well. The relation between Grüneisen parameter and Poisson's ratio is as follows: $\gamma = \frac{3}{2}\frac{1+\nu}{2-3\nu}$ [75]. The calculated values of Grüneisen parameters at different pressures are presented in Table 7.

*Melting temperature*: Information on the melting temperature of a compound is very important for practical applications at different temperatures. High melting temperature of a compound has lower thermal expansion and high binding energy and vice versa. We have calculated the melting temperature $T_m$ using the following equation [76]:

$$T_m = 354 + \frac{4.5(2C_{11} + C_{33})}{3} \tag{13}$$

The calculated values of melting temperatures of two phases of $P2_1/m$ and $Pmn2_1$ MgVH$_6$ are listed in Table 7. The melting temperatures of $Pmn2_1$ are very high at different pressures. This also agrees with the previously estimated bulk modulus, Debye temperature, and hardness. The melting temperature increases with the increase in the pressure for both the structures.

*Thermal expansion coefficient*: The thermal expansion coefficient (TEC) of a material is connected to many other physical properties, such as thermal conductivity, heat capacity, temperature variation of the energy band gap and electron effective mass. The thermal expansion coefficient of a material can be obtained using the following equation [67]:

$$\alpha = \frac{1.6 \times 10^{-3}}{G} \tag{14}$$

The relation between thermal expansion coefficient and the melting temperature can be approximated as $\alpha \approx 0.02/T_m$ [76]. From Table 7 it is observed that for $Pmn2_1$ has lower thermal expansion compared to $P2_1/m$.

Pressure variation of Debye temperature and the Grüneisen parameter are shown in Fig. 9.

Table 7. Grüneisen parameter γ, Debye temperature $\theta_D$ (K), thermal expansion coefficient $\alpha$ (K$^{-1}$), minimum thermal conductivity (W/m.K), and melting temperature $T_m$ (K) of MgVH$_6$ in the monoclinic and orthorhombic structures.

| Space group | Crystal structure | Pressure P (GPa) | γ | $\theta_D$ | $\alpha$ (10$^{-5}$) | $k_{min}$ | $T_m$ | Ref. |
|---|---|---|---|---|---|---|---|---|
| $P2_1/m$ | Monoclinic | 0 | 1.64 | 676.25 | 4.24 | 1.01 | 807.14 | [This] |
| | | 5 | 1.78 | 728.56 | 3.60 | 1.11 | 938.85 | |
| | | 10 | 1.79 | 764.01 | 3.22 | 1.18 | 1070.70 | |
| | | 15 | 1.98 | 790.70 | 2.99 | 1.24 | 1178.11 | |
| $Pmn2_1$ | Orthorhombic | 100 | 1.09 | 2330.44 | 0.41 | 9.18 | 2906.54 | [This] |
| | | 125 | 1.99 | 1606.44 | 0.88 | 6.45 | 3200.57 | |
| | | 150 | 2.35 | 1464.29 | 1.06 | 5.97 | 3778.52 | |
| | | 175 | 2.29 | 1585.24 | 0.89 | 6.56 | 4214.74 | |
| | | 200 | 2.28 | 1667.94 | 0.79 | 7.00 | 4646.19 | |

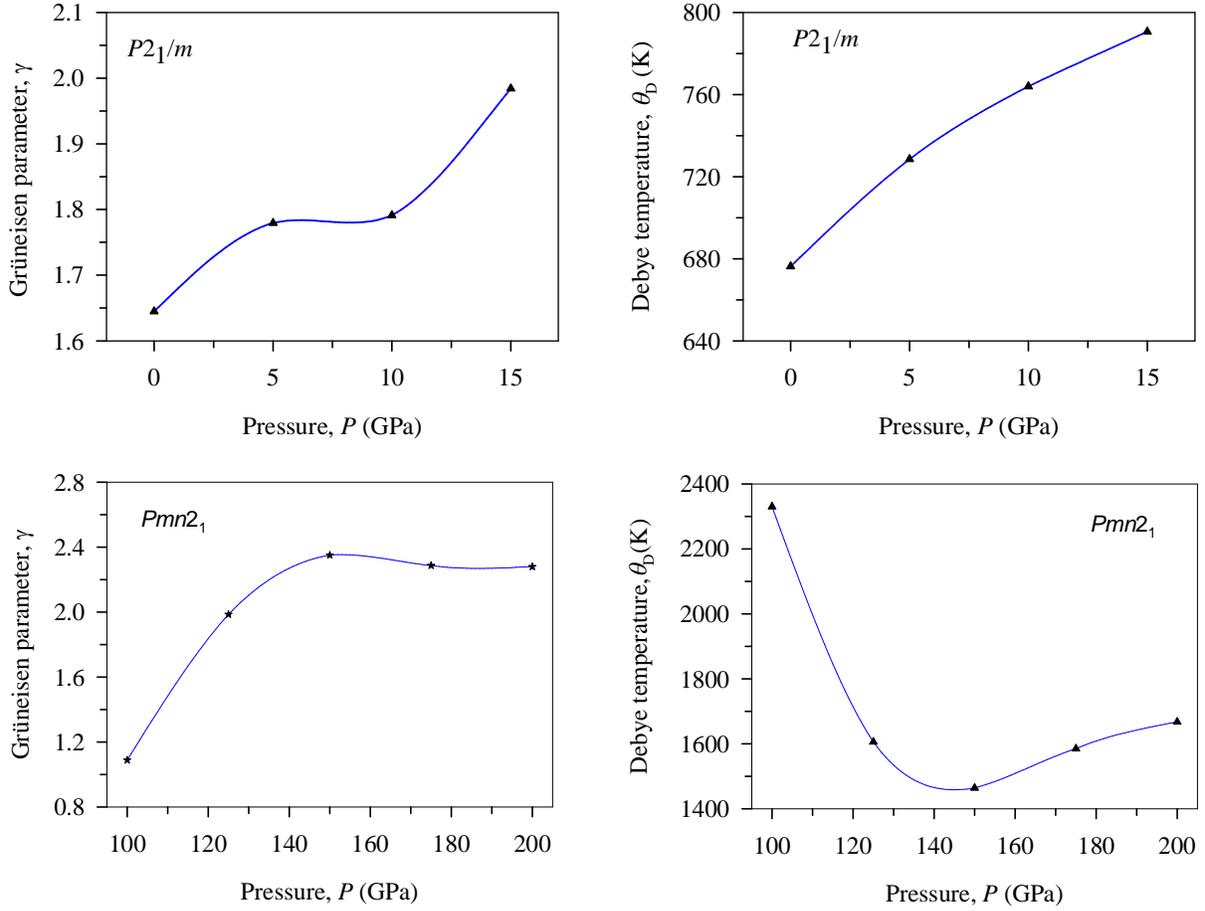

Figure 9: Grüneisen parameter and Debye temperature of MgVH$_6$ at different pressures.

*3.7 Superconducting properties*

Zheng et al. [36] predicted superconductivity in the orthorhombic phase of MgVH$_6$ under pressure. They explored the electronic band structure, phonon spectrum, the Eliashberg function and superconducting transition temperature. The electron-phonon coupling constant was also estimated.

We have revisited the electronic band structures (not shown in this paper) of MgVH$_6$ in the orthorhombic and monoclinic structures over a wider pressure range. As far as superconductivity is concerned, the electronic energy density of states (DOS) at the Fermi level and the repulsive Coulomb pseudopotential are two important parameters [36,77-81]. It should be noted that MgVH$_6$ in the monoclinic phase is not superconducting [36]. Table 8 discloses the computed parameters related to superconductivity. For completeness, the electronic energy density of states at the Fermi level and the repulsive Coulomb pseudopotential of non-superconducting MgVH$_6$ in the monoclinic phase are also included in Table 8.

Table 8. DOS at the Fermi level N($E_F$) (States/eV), repulsive Coulomb pseudopotential $\mu^*$, electron-phonon coupling constant $\lambda$, and $T_c$ (K).

| Space group | Crystal structure | Pressure P (GPa) | N($E_F$) | $\mu^*$ | $\lambda$ | $T_c$ | Ref. |
|---|---|---|---|---|---|---|---|
| $P2_1/m$ | Monoclinic | 0 | 5.00 | 0.217 | - | - | [This] |
| | | 5 | 4.52 | 0.213 | - | - | |
| | | 10 | 4.19 | 0.210 | - | - | |
| | | 15 | 3.96 | 0.208 | - | - | |
| $Pmn2_1$ | Orthorhombic | 100 | 1.78 | 0.166 | 1.06 | 104.7 | [This] |
| | | 125 | 1.50 | 0.156 | 0.89 | 53.8 | |
| | | 150 | 1.27 | 0.145 | 0.75 | 34.9 | |
| | | | 2.40 | 0.10, 0.13 | 1.43 | 27.6, 25.0 | [36] |
| | | 175 | 1.13 | 0.138 | 0.67 | 28.6 | [This] |
| | | 200 | 1.07 | 0.134 | 0.63 | 26.1 | |

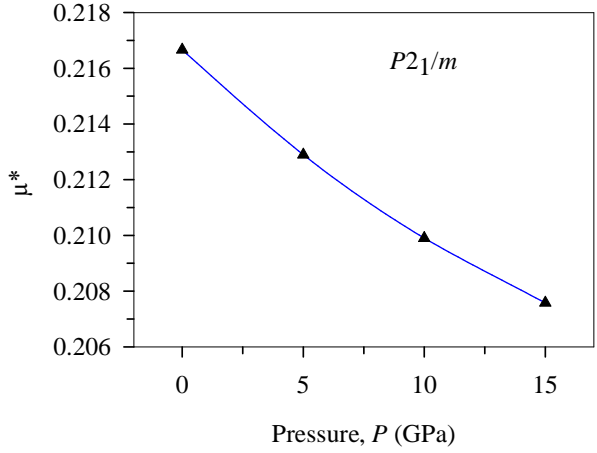
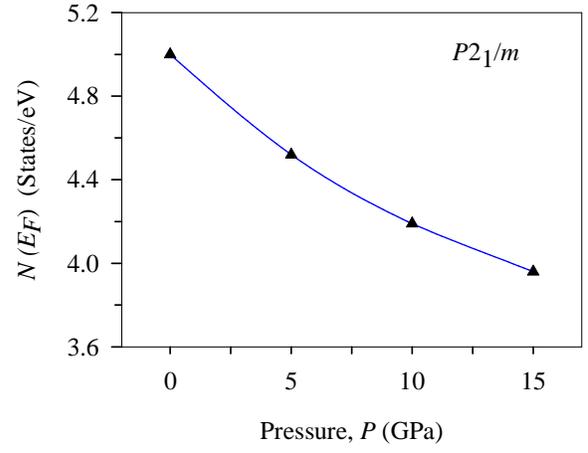
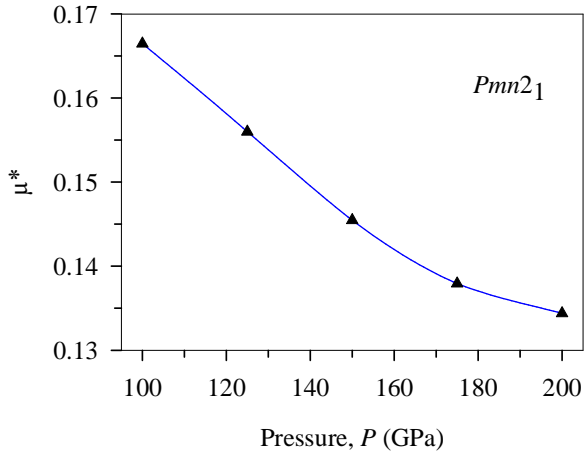
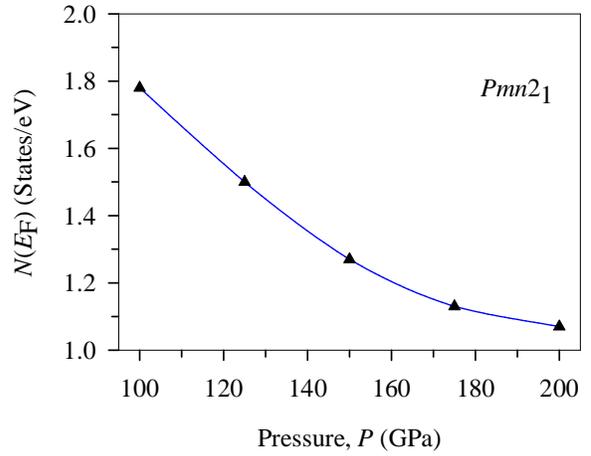

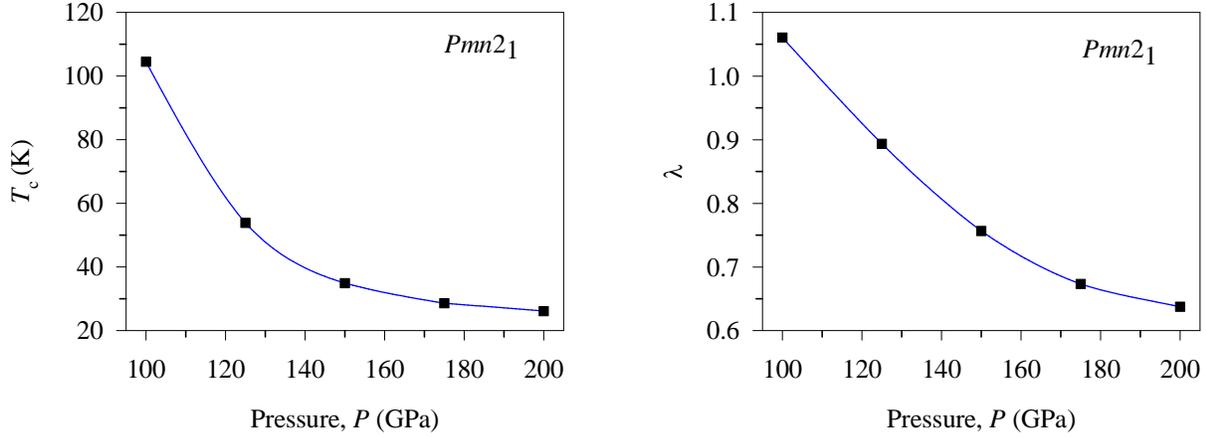

Figure 10: The repulsive Coulomb pseudopotential $\mu^*$, electronic density of states at the Fermi level $N(E_F)$ (States/eV) (in both the phases), electron phonon coupling constant $\lambda$, and superconducting critical temperature $T_c$ of $MgVH_6$ at different pressures in the orthorhombic structure only.

The McMillan equation given below is used to estimate $T_c$ [82]:

$$T_C = \frac{\theta_D}{1.45} \exp\left[-\frac{1.04(1+\lambda)}{\lambda - \mu^*(1+0.62\lambda)}\right] \quad (15)$$

In Eqn. (15), $\lambda$ is the electron-phonon coupling constant and $\mu^*$ is the repulsive Coulomb pseudopotential. The pressure variation in the electron-phonon coupling constant has been estimated following the procedure adopted in Ref. [78]. We have used the computed value of the $\lambda$ from the Eliashberg spectral function at 150 GPa [36] as the reference value. In order to evaluate the pressure dependence of $T_c$, first, we need the pressure dependent variations of $\theta_D$ and $N(E_F)$ (since $\lambda = N(E_F)V_{e-ph}$, here $V_{e-ph}$ is the electron-phonon coupling potential; a term which is a weakly varying function of pressure in most cases). The variations of $N(E_F)$, $\mu^*$, $\lambda$, and $T_c$ with pressure are shown in Fig. 10. The variations of $\theta_D$ as a function of pressure have been displayed in Fig. 9. In our calculations, the value of Coulomb pseudopotential, $\mu^*$ has been determined using the following equation [83]:

$$\mu^* = \frac{0.26\, N(E_F)}{1 + N(E_F)} \quad (16)$$

The repulsive Coulomb pseudopotential is detrimental to the formation of Cooper pairs which is essential for superconductivity. High values of $\lambda$, $\theta_D$, and low value of $\mu^*$ favor high-temperature superconductivity. From Table 8 it is observed that $T_c$ decreases under pressure primarily due to the decrease in the DOS at the Fermi level. The DOS at the Fermi level is the dominant parameter known to affect $T_c$ via electron-phonon coupling constant [77]. Our calculations, therefore, suggest that the change in the DOS at the Fermi level under pressure is the origin of the pressure dependence variations of $T_c$. Very large value of $T_c$ of $MgVH_6$ at 100 GPa in the orthorhombic structure results from an anomalously large value of the Debye temperature and relatively high value of the electronic energy density of states at the Fermi level. It is interesting to note that the DOS at the Fermi level is significantly higher in the non-superconducting monoclinic $MgVH_6$. At the same time, $MgVH_6$ in this structure has significantly lower Debye temperature and higher repulsive Coulomb pseudopotential compared to those in the orthorhombic structure.

*3.8 Optical properties*

The optical functions of $MgVH_6$ are evaluated for the two polarization directions [100] and [001] of incident photons up to 30 eV at different pressures. The overall pressure dependence is weak. In this section we show the representative results for 0 GPa for the $P2_1/m$ and for 100 GPa for the $Pmn2_1$ structures. The frequency/energy dependent dielectric function can be written as follows:

$$\varepsilon(\omega) = \varepsilon_1(\omega) + \varepsilon_2(\omega) \qquad (17)$$

Here, $\omega$ is the angular frequency of the electromagnetic wave, and $\varepsilon_1(\omega)$ and $\varepsilon_2(\omega)$ are the real and imaginary parts of the dielectric functions. The imaginary part of dielectric functions can be expressed as follows [49,84]:

$$\varepsilon_2(\omega) = \frac{2\pi e^2}{\Omega \varepsilon_0} \sum_{k,v,c} |\psi_k^c| \boldsymbol{u} \cdot \boldsymbol{r} |\psi_k^v|^2 \delta(E_k^c - E_k^v - E) \qquad (18)$$

Here, $\omega$ is the phonon angular frequency, $e$ is the electronic charge, $\Omega$ is the unit cell volume, **u** is the unit vector along the polarization of the incident electric field and $\psi_k^c$ and $\psi_k^v$ are the wave functions for conduction and valence band electrons at a particular value of *k*, respectively. In metallic solids the imaginary part of the dielectric constant comes from two contributions; inter-band and intra-band optical transitions. $MgVH_6$ in both the structures are metallic and, therefore, the intra-band transitions have crucial impact at the far infrared regions i.e., low energy part of the electromagnetic spectrum. To take account of the intraband transition for metallic material, the Drude damping correction is required [85,86]. Optical properties of $MgVH_6$ have been calculated considering screened plasma energy of 10 eV and a Drude damping of 0.08 eV.

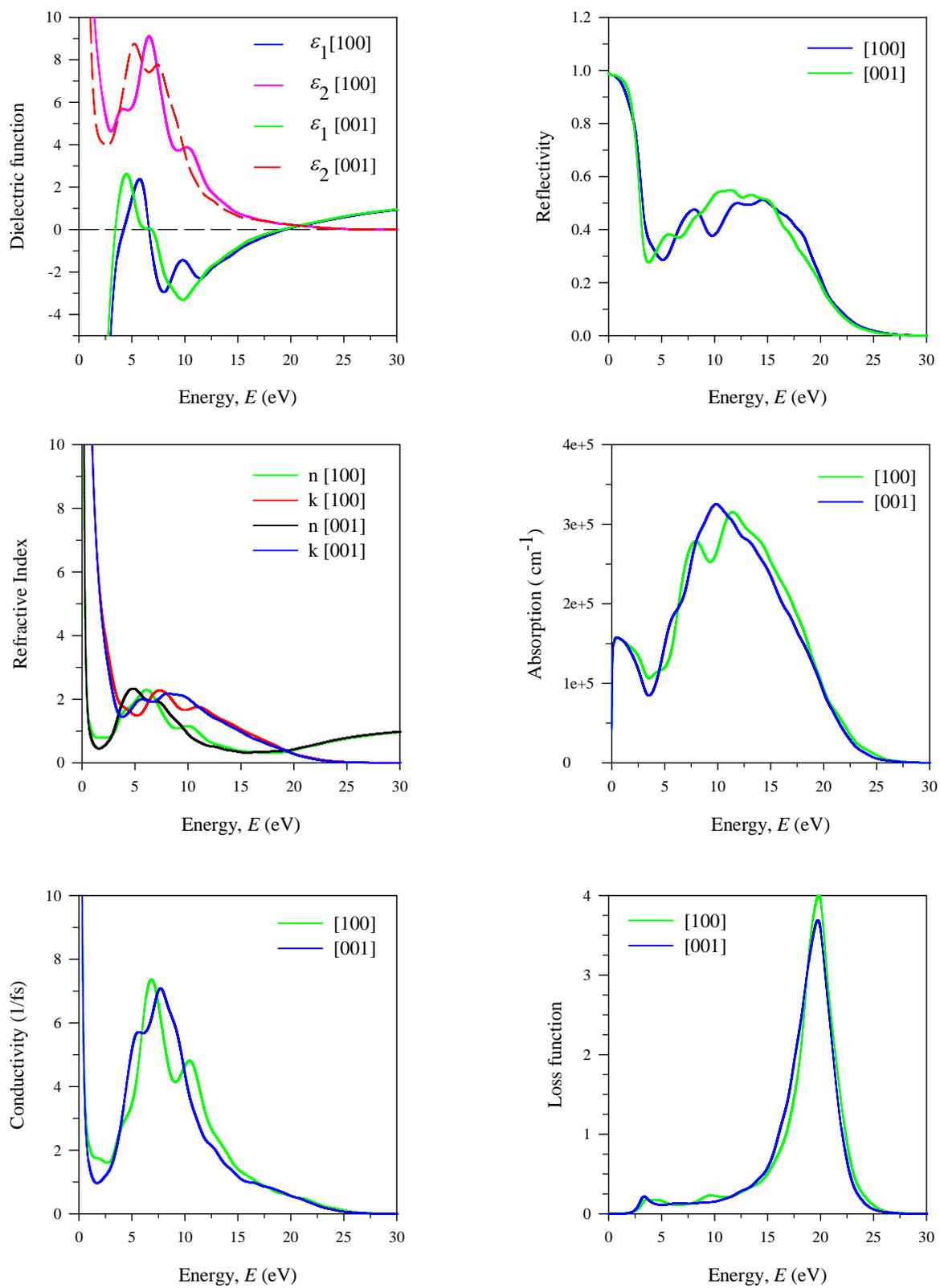

Figure 11: Optical properties of ($P2_1/m$) MgVH$_6$ at 0 GPa.

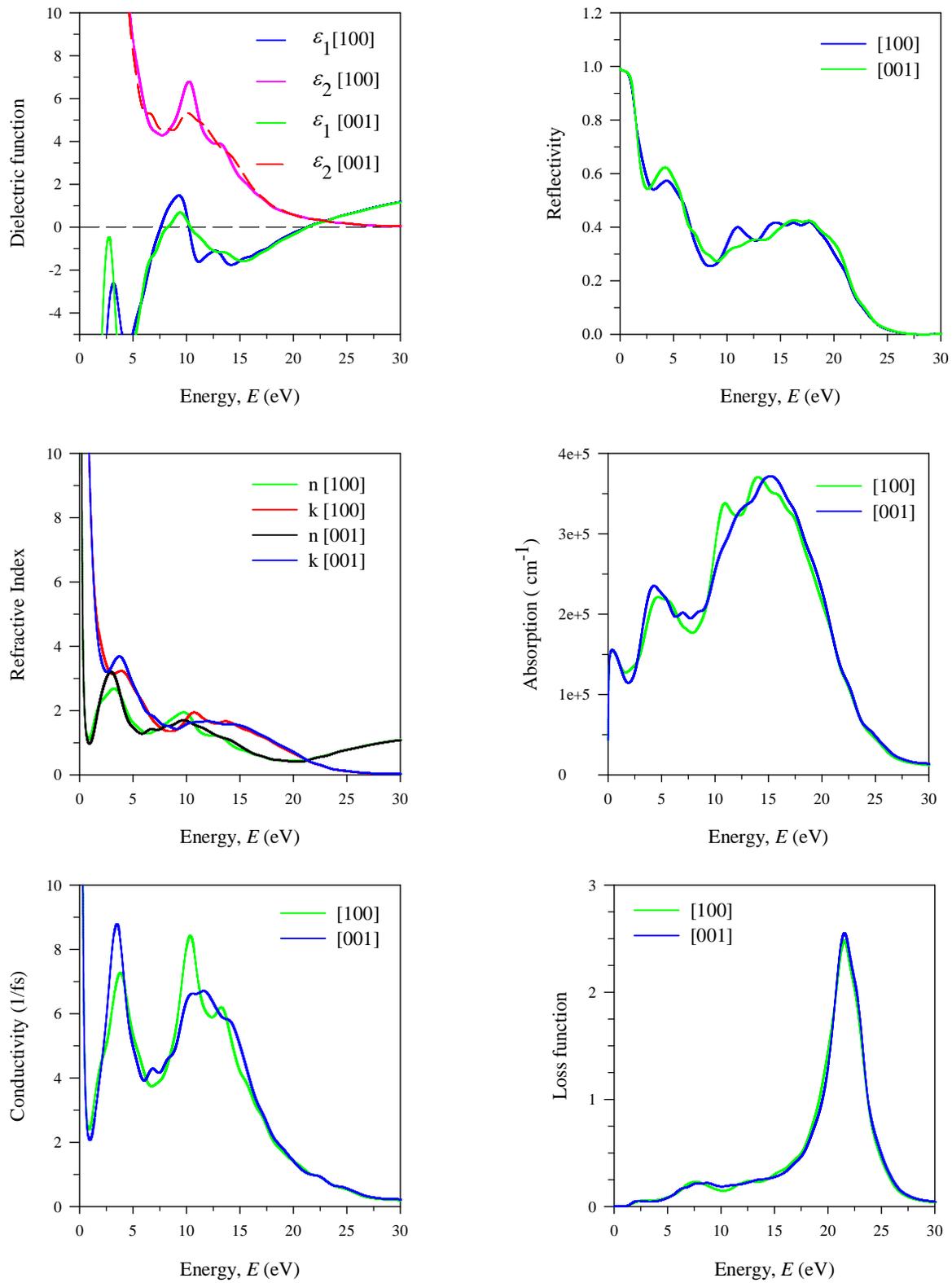

Figure 12: Optical properties of (*Pmn*2$_1$) MgVH$_6$ at 100 GPa.

Both real and imaginary parts of the dielectric function in monoclinic and orthorhombic structures show metallic character. The reflectivity is very high in the infra-red and visible regions of the electromagnetic spectra. According to Li et al. [85] a compound will be capable of reducing solar heating if it has a reflectivity ~44% in the visible light region. Therefore, $MgVH_6$ is a promising candidate for the practical usage as a coating material to avoid solar heating. The real part of the complex refractive index which determines the phase velocity of electromagnetic waves in $MgVH_6$, is low in the visible region. The imaginary part, known as the extinction coefficient which measures the attenuation is high in the visible region. Both optical conductivity and absorption coefficient show metallic features. $MgVH_6$ is an efficient absorber of ultraviolet radiation. Loss function is a useful optical parameter describing the energy loss of a fast electron traveling in a material. The energy at which the loss function is a maximum is known as the plasma energy. The plasmons are excited and the reflectivity, absorption coefficient, and photoconductivity fall sharply at this energy. The material shows insulating behavior in response to incident photons above the plasma energy. The loss peaks are observed for $P2_1/m$ structure at 19.812 eV and 19.68 eV for the [100] and [001] electric field polarizations, respectively. For the $Pmn2_1$ structure, the loss peaks appear at the same energy, at 21.50 eV, for the [100] and [001] electric field polarizations. The optical anisotropy is low for $MgVH_6$ in both the structures.

## 4. Conclusions

DFT-based first-principles calculations have been carried out to find out the effects of hydrostatic pressure on the structural, elastic, thermophysical, superconducting state, and optical properties of $MgVH_6$ in the $P2_1/m$ and $Pmn2_1$ structures. The lattice parameters and unit cell volume decrease with pressure. The variation is systematic in the monoclinic phase but at high pressures the *b*- and *c*-axis lattice parameters show opposing trends. Both $P2_1/m$ and $Pmn2_1$ phases of $MgVH_6$ are thermodynamically stable. On the other hand, $P2_1/m$ phase of $MgVH_6$ is mechanically stable under the pressures considered here but the $Pmn2_1$ phase of $MgVH_6$ was found to be mechanically unstable under the pressures considered. Moreover, pressure variations of different elastic and thermophysical parameters indicate towards structural instability of $MgVH_6$ in the orthorhombic phase close to 100 GPa. For the $P2_1/m$ structure, both Poisson's ratio and Pugh's ratio are consistent with ductility. On the other hand, for the $Pmn2_1$ structure, the compound is predicted to be brittle at 100 GPa, and from 125 GPa – 200 GPa, it shows ductility. $MgVH_6$ in both the structures is elastically anisotropic. The hardness of $MgVH_6$ in the orthorhombic structure is very high at 100 GPa and falls sharply for further increase in pressure. The orthorhombic phase ($Pmn2_1$) has lower thermal expansion coefficient compared to that of monoclinic phase ($P2_1/m$). This also agrees with the elastic moduli, Debye temperature, and hardness values. The superconducting transition temperature decreases systematically with increasing pressure from a high value of 104.7 K at 100 GPa in the orthorhombic structure. This implies that a higher $T_c$ might be achievable in the orthorhombic $MgVH_6$; the prime challenge here would be to ensure structural stability. The optical parameters are investigated in details.

Both the phases reveal clear metallic characters, high absorption capability of ultraviolet radiation and high reflectivity of infrared and visible light. Optical anisotropy is low in orthorhombic and monoclinic MgVH$_6$.

**Acknowledgements**

S. H. N. acknowledges the research grant (1151/5/52/RU/Science-07/19-20) from the Faculty of Science, University of Rajshahi, Bangladesh, which partly supported this work.

**Data availability**

The data sets generated and/or analyzed in this study are available from the corresponding author on reasonable request.

**Declaration of interest**

The authors declare that they have no known competing financial interests or personal relationships that could have appeared to influence the work reported in this paper.

[63] Ravindran, P., Fast, L., Korzhavyi, P. A., Johansson, B., Wills, J., & Eriksson, O. (1998). Density functional theory for calculation of elastic properties of orthorhombic crystals: Application to TiSi$_2$. *Journal of Applied Physics*, *84*(9), 4891-4904.

[64] Zhu, S., Zhang, X., Chen, J., Liu, C., Li, D., Yu, H., & Wang, F. (2019). Insight into the elastic, electronic properties, anisotropy in elasticity of manganese borides. *Vacuum*, *165*, 118-126.

[65] Ranganathan, S. I., & Ostoja-Starzewski, M. (2008). Universal elastic anisotropy index. *Physical Review Letters*, *101*(5), 055504.

[66] Schreiber, E., Anderson, O. L., Soga, N., & Bell, J. F. (1975). Elastic Constants and Their Measurement. *Journal of Applied Mechanics*, *42*(3), 747.

[67] Ali, M. A., Hossain, M. M., Islam, A. K. M. A., & Naqib, S. H. (2021). Ternary boride Hf$_3$PB$_4$: Insights into the physical properties of the hardest possible boride MAX phase. *Journal of Alloys and Compounds*, *857*, 158264.

[68] Rano, B. R., Syed, I. M., & Naqib, S. H. (2020). Ab initio approach to the elastic, electronic, and optical properties of MoTe$_2$ topological Weyl semimetal. *Journal of Alloys and compounds*, *829*, 154522.

[69] Teter, D. M. (1998). Computational alchemy: The search for new superhard materials. *MRS Bulletin*, *23*(1), 22-27.

[70] Tian, Y., Xu, B., & Zhao, Z. (2012). Microscopic theory of hardness and design of novel superhard crystals. *International Journal of Refractory Metals and Hard Materials*, *33*, 93-106.

[71] Chen, X. Q., Niu, H., Li, D., & Li, Y. (2011). Modeling hardness of polycrystalline materials and bulk metallic glasses. *Intermetallics*, *19*(9), 1275-1281.

[72] El-Adawy, A., & El-KheshKhany, N. (2006). Effect of rare earth (Pr$_2$O$_3$, Nd$_2$O$_3$, Sm$_2$O$_3$, Eu$_2$O$_3$, Gd$_2$O$_3$ and Er$_2$O$_3$) on the acoustic properties of glass belonging to bismuth–borate system. *Solid State Communications*, *139*(3), 108-113.

[73] Anderson, O. L. (1963). A simplified method for calculating the Debye temperature from elastic constants. *Journal of Physics and Chemistry of Solids*, *24*(7), 909-917.

[74] Clarke, D. R. (2003). Materials selection guidelines for low thermal conductivity thermal barrier coatings. *Surface and Coatings Technology*, *163*, 67-74.

[75] Mirzai, A., Ahadi, A., Melin, S., & Olsson, P. A. (2021). First-principle investigation of doping effects on mechanical and thermodynamic properties of Y$_2$SiO$_5$. *Mechanics of Materials*, *154*, 103739.

[76] Fine, M. E., Brown, L. D., & Marcus, H. L. (1984). Elastic constants versus melting temperature in metals. *Scripta Metallurgica*, *18*(9), 951-956.

**CRediT author statement**

**Md. Ashraful Alam:** Methodology, Software, Formal analysis, Writing-Original draft. **F. Parvin:** Supervision, Writing-Reviewing and Editing. **S. H. Naqib:** Conceptualization, Supervision, Formal analysis, Writing- Reviewing and Editing.